\documentclass{sigchi}

\usepackage{balance}  
\usepackage{graphics} 
\usepackage{graphicx}
\usepackage[english]{babel}
\usepackage[T1]{fontenc}
\usepackage{txfonts}
\usepackage{mathptmx}
\usepackage[pdftex]{hyperref}
\usepackage{color}
\usepackage{booktabs}
\usepackage{textcomp}
\usepackage{dblfloatfix}

\usepackage[utf8]{inputenc}

\usepackage{array}
\newcolumntype{L}[1]{>{\raggedright\let\newline\\\arraybackslash\hspace{0pt}}m{#1}}

\usepackage{microtype} 
\usepackage{ccicons}  

\usepackage{todonotes}
  
\usepackage{varwidth}

\def\plaintitle{A Common Framework for Audience Interactivity}

\def\emptyauthor{}
\def\plainkeywords{Performance interaction; Audience Interactivity; Transmedia; Immersive Experiences; Media arts}

\makeatletter
\def\url@leostyle{%
  \@ifundefined{selectfont}{
    \def\UrlFont{\sf}
  }{
    \def\UrlFont{\small\bf\ttfamily}
  }}
\makeatother
\def\pprw{8.5in}
\def\pprh{11in}

\definecolor{linkColor}{RGB}{6,125,233}
\hypersetup{%
  pdftitle={\plaintitle},
  pdfauthor={\emptyauthor},
  pdfkeywords={\plainkeywords},
  bookmarksnumbered,
  pdfstartview={FitH},
  colorlinks,
  citecolor=black,
  filecolor=black,
  linkcolor=black,
  urlcolor=linkColor,
  breaklinks=true,
}

\begin{document}

\title{\plaintitle}


\numberofauthors{3}
\author{%
  \alignauthor{Alina Striner\\
    \affaddr{University of Maryland}\\
    \affaddr{College Park, USA}\\
    \email{algol001@umd.edu}}\\
  \alignauthor{Sasha Azad\\
    \affaddr{North Carolina State University}\\
    \affaddr{Raleigh, USA}\\
    \email{sazad@ncsu.edu}}\\
  \alignauthor{Chris Martens\\
    \affaddr{NC State University}\\
    \affaddr{Raleigh, USA}\\
    \email{martens@csc.ncsu.edu }}\\
}

\maketitle

\begin{abstract}
Audience interactivity is interpreted differently across domains. This research develops a framework to describe audience interactivity across a broad range of experiences. We build on early work characterizing child audience interactivity experiences, expanding on these findings with an extensive review of literature in theater, games, and theme parks, paired with expert interviews in those domains. The framework scaffolds interactivity as nested \textit{spheres of audience influence}, and comprises a series of dimensions of audience interactivity including a \textit{Spectrum of Audience Interactivity}. This framework aims to develop a common taxonomy for researchers and practitioners working with audience interactivity experiences.
\end{abstract}

\category{H.5.m. Information interfaces and presentation (e.g.,
HCI): Miscellaneous.}{}{}


\section{Introduction}
Interactivity has the power to immerse and empower audiences across various genres and mediums, including theater, music, film, theme parks, and literary entertainment.
These mediums differ in terminology, sometimes describing interactive approaches as {\em participatory} or {\em immersive}, but generally refer to a desired outcome of more fulfilling storytelling experiences.
%

Narratives and storytelling perpetually evolve and new interactive opportunities for the audience of these narratives have emerged in each medium over time. In \textit{Hamlet and the Holodeck}, Murray takes a detailed look at storytelling in the digital age, specifically narratives that dealt with alternative storylines or required audience participation. She argues that, in the future, fiction authors would be challenged to create rules for the space of interaction, rather than the narrative itself. These rules would transform the interactor from a merely receptive reader and would instead promote an immersive and reactive storytelling experience~\cite{murray2017hamlet}.

Designing interactive experiences often means learning from previous work and building experiences using available tools. Since audience interactivity experiences exist in a range of domains and contexts, designers are often limited to learning from their area of expertise. To develop new forms of artistic expression, HCI practitioners require a common framework and taxonomy to compare and learn from diverse experiences.

Interactivity literature suggests that several dimensions of interactivity characterize the breadth of interactive experiences. For instance, Macintyre~\cite{macintyre2002three} describes audience experiencing a courtroom narrative firsthand using augmented reality, Ito~\cite{ito2005intertextual} illustrates audience members ability to create personalized narratives interacting with Disney characters, and McAllister~\cite{mcallister2004interactive} details audience's ability to edit music performance compositions. Several rudimentary existing models~\cite{everett1986communication,rafaeli1988new,benford2006frame,zimmerman2004narrative,steuer1992defining} have describe a few dimensions of audience interactivity, often lumping the audience’s experience under the umbrella term, `interactive,' however literature suggests that more complicated relationships need to be defined that address Murray’s fully interactive world~\cite{murray2017hamlet}. 

HCI practitioners also needs to explicitly consider the appropriateness of technology to support design goals. 
in pursuit of the dream of the Holodeck,
HCI often produces novel technology and techniques to facilitate more immersive modes of interaction, such as virtual reality and gamification. While compelling, novel technology fads may compromise practitioner design goals in favor of technology affordances. As well detracting from design goals, technologies incongruent with design goals may stall adoption of new technology. For instance, a number of entertainment domains have experimented with virtual reality~\cite{rec_room_VRgame,curtis2016_makingPearl,de2010immersive_Journalism} following hearsay about its immersive power, yet the medium has yet to find a mainstream audience, in part because it's been over-appropriated into experiences not best suited for the medium.~\cite{forbesVR}. 

This work develops a framework and taxonomy to describe audience interactivity across a broad range of domains and experiences in order to help researchers and practitioners juxtapose and learn from divergent audience experiences. We build on early work that developed a spectrum of audience interactivity for children, and expand on these findings with an extensive review of literature in theater, games, and theme parks, paired with expert interviews in those domains. Our framework uses nested \textit{spheres of audience influence} and a series of dimensions to describe audience interactivity, including a new \textit{Spectrum of Audience Interactivity}. This framework aims to be a common taxonomy for researchers and practitioners working with audience interactivity experiences; it can help practitioners consider challenges inherent to different types of audience interactivity design, and allow artists to develop new interactive media without succumbing to fads.

\section{Background}
In this section we discuss the phenomenon we wish to taxonomize, which is how storytelling has evolved to incorporate audience interaction, resulting in more immersive and engaging experiences. We define these terms along the way.
\subsection{Engagement and Immersion}
A primary goal for entertainment domains is to engage and immerse audiences~\cite{pine1999experience,pine2013experience}. Engagement refers to intensity of involvement that provides an immensely compelling and rewarding experience~\cite{garris2002games}. Paired with engagement, immersion refers to a sense of being surrounded by another reality that takes over a person's attention and perception \cite{ermi2005fundamental,cummings2016immersive}, which  enhances motivation and creates a feeling of `deep play' that fosters emotional investment~\cite{Tynansylvester2013designing}. Presence, a `transportation' effect created through sensory immersion~\cite{mcmahan2003immersion}, has been described as another facet of engagement and immersion. 

Several constructs~\cite{goldman2014predicting} have been proposed to describe engagement and immersion, including presence and flow. Brockmyer~\cite{brockmyer2009development} and Cummings~\cite{cummings2016immersive} suggest that engagement is often created through a sense of presence, the sense of `being there,' surrounded by another reality that takes over a person's attention and perception~\cite{ermi2005fundamental,cummings2016immersive}. 
Ermi and Mayra~\cite{ermi2005fundamental} present a multidimensional model of immersion consisting of sensory immersion, overpowering sensory information through large screens and powerful sounds, challenge-based immersion, a balance motor or mental skills and abilities, and imaginative immersion, absorption within a fictional narrative and world. McMahan~\cite{mcmahan2003immersion} explains that presence closely relates to the phenomenon of distal attribution, the ability to reference perceptions of an externalization space beyond the limits of the sensory organs. Since we perceive the world using multiple senses, immersion is often associated with the high pervasiveness and fidelity of multisensory inputs, such visual, auditory, olfactory and tactile cues~\cite{shernoff2009cultivating}. The pinnacle of challenge-based immersion has been described as {\em flow} ~\cite{csikszentmihalyi1990flow,engeserRheinberg2008flow,kiili2014flow,shernoff2009cultivating}, a state of total absorption in a task, which produces confidence and self-esteem and yields optimal performance~\cite{ermi2005fundamental}.

Entertainment literature embodies the importance of engagement and immersion. For instance, Green et al. have found that narrative transportation can affect persuasion and belief change, as well as enjoyment~\cite{green2004understanding}.
Likewise, motivation and decision research demonstrates that effectively integrated  judgment, behavior, and feedback cycles interwoven with engagement can lead to  increased confidence, persistence, and effort~\cite{garris2002games} for novice learners. 

\subsection{Fantasy and Narratives}
Throughout history, fantasy and narratives have fulfilled audience engagement and immersion needs by transporting audiences, constructing experiences with ``\textit{cognitive, emotional, and imagery involvemen}t'' ~\cite{green2004understanding} that help them make sense of the world. Theme parks fulfill engagement and immersion needs by constructing fantasies of another place and another time~\cite{milman200713, cross2005playful}. Designed purposely to be isolated, theme parks invite guests to leave the real world at the parking lot and gain temporary ``citizenship'' to a fantasy world~\cite{bukatman1991there}~\cite{durrant2012pursuing}.
These fantasy worlds escape from the rules and conventions of the outside world~\cite{van1991smile}; there are often no clocks~\cite{cross2005playful} nor defined social barriers between guests ~\cite{bradford2015domesticating}.

Building on Miller et al.~\cite{miller1990critical}, Zimmerman ~\cite{zimmerman2004narrative} defines narrative as having an initial state, a change in that state, and some insight brought about by that change in state. Narratives further fantasy by transporting audiences into the experiences of characters ~\cite{salen2004rules}. For instance, games explicitly offer players a role in the story~\cite{mateas2006interaction}, such as the interactive drama Fa\c{c}ade~\cite{mateas2003faccade}, which demonstrates virtual characters responding to a player-performer. Even allegedly `passive' narratives have a powerful ability to {\em transport} audiences~\cite{green2004understanding}, creating ``an experience of cognitive [and] emotional involvement.'' 

\subsection{Audience Interaction}

The capability of the audience to alter and transform experiences has been considered on the one hand the empowerment of audience \cite{mcmillan2002four}, and on the other the dissolution of the traditional idea of an audiencehood \cite{brooker2003conclusion}. With the rise of interactive audiences, the role of the audience in a performance has changed. Based upon empirical data provided by questionnaires answered by 6700 players, Yee~\cite{yee2002facets} added an `immersionist' factor of audience response to Bartle's original scheme~\cite{bartle2004designing}. Immersion represents the desire of the participant to detach from real life and incorporate themselves into a fantasy world, enjoy wandering and exploring, role-playing their characters, or using their characters to try out new personalities and being part of a ongoing story. 

Bartle's later work describes the degree of immersion of the audience member with their character ranging from the avatar at one extreme to a persona at another. In audience-driven interactive theater performances, 
authors invite audiences into an immersionist world, to interact with the narrative performed as a group, changing the direction of the story. Early examples include {\em The Night of January 16th}~\cite{rand1971night}, in which audience members play the role of a jury in a court room, and {\em Drood}~\cite{pointer1996charles}, a musical adaptation of a murder mystery in which the audience assists with solving the mystery. 

While previous work such as Bartle's has described the varying degrees of audience immersion in a narrative, the levels of interaction between the triad of immersionist, audience, and performers are yet to be thoroughly explored. With this paper, we analyze the interactions between all three parties and present a taxonomy that would enable the designers of immersive experiences to consider the range of audience interactivity available to them, ranging from passive attention to a more active performance.

\



\section{Previous Efforts To Characterize Audience Interactivity}

Previous research has endeavored to characterize interactivity in media experiences. Relatively simple models include Everette's single dimensional scale rating the interactivity of communication technologies \cite{everett1986communication}, and Rafaeli \cite{rafaeli1988new}, who classifies media based on responsiveness to audiences. Working with textual narratives, Zimmerman~\cite{zimmerman2004narrative} identifies four modes of audience interactivity that complement our goal of broadly describing a  taxonomy of interactivity for transmedia; cognitive interactivity, a response to and internalization of a narrative, Functional Interactivity---interactions with the physical reality and materiality of a text, such as turning pages; Explicit Interactivity---participation in the narrative flow, making choices, participating in narrative events; and Meta-interactivity--interaction outside the narrative, which includes creating, deconstructing, and reconstructing a narrative.

Multiple models of interactivity characterize interactivity by the choices available to the audiences, and how audiences exercise those choices ~\cite{goertz1995interaktiv,laurel1986interface,laurel1997interface,laurel1991computers}. Steuer expands on Everette's characterization of interactivity with a two dimensional model based on vividness, the ability of a medium to provide a rich mediated environment, and interactivity, the ability of the user to modify vividness of their experience \cite{steuer1992defining}. While Steuer's method is a highly cited as a measure of immersion and engagement, it notably fails to provide explicit criteria for new experiences to be mapped on to his scale\cite{jensen1998interactivity}. Brenda Laurel's three dimensional model further characterizes interactivity by frequency of interactivity, the range of choices available, and the extent to which choices affect experience \cite{laurel1991computers}, and Goertz introduces a four dimension scale describing interactivity through linearity and degrees, numbers, and modifiability of choice~\cite{goertz1995interaktiv,jensen1998interactivity}. 

\subsection{A Unified Framework to Describe Audience Interactivity} 

Previous work by Striner ~\cite{Striner2017_Transitioning} stewarded a first step toward understanding the many ways in which technology can allow audience members to interact with performance. This work organically developed a spectrum of interactivity from children's codesign sessions using Cooperative Inquiry (CI) derived from \textit{Participatory Design}~\cite{druin1999cooperative,guha2013cooperative}. 



In contrast to existing models that narrowly characterize interactivity, we build on this spectrum of interactivity to create a common framework and vocabulary for audience interactivity that characterizes audience interactivity across multiple domains using several parallel dimensions. The framework further describes the range of audience interactivity through an audience member's influence over their own experience, that of other audience members, performers, and in the larger performance. Using this framework, our contribution creates a way for designers to compare performance experiences and identify gaps and challenges in interactivity.

\section{Goals}

The overarching goal of this work is develop a common framework for audience interactivity by building on prior work by Striner ~\cite{Striner2017_Transitioning} that proposed a spectrum of audience interactivity for children in the musical performances domain. This work iterates on this framework based on insights from a comprehensive review of literature across three diverse domains, theater, games, and theme parks. We pair this review with expert interviews from those domains to understand practitioner insight into audience interactivity. Based on the literature review and expert interviews, we propose a common framework to describe audience interactivity across domains. 

This research:
\begin{enumerate}
\item Performs a review of interactivity literature across the three domains, theatre, games and theme parks
\item Conducts in-depth expert interviews with domain experts to validate themes
\item Describes a new framework for audience interactivity based on literature and expert interview themes 
\end{enumerate}


\subsection{Domain Choice} 

Audience interactivity exists in a broad range of entertainment and education domains,~\cite{WarholDance_gahr_2017,ARJournalism_pavlik2013emergence,shipmanblending,gilbert1998building_InteractivityEducation}. Rather than trying to tackle all the possible literature that exists across domains, we focused on search on three primary domains: theater performances, games, and theme parks. These domains were chosen as representative of diverse audiences, interaction modalities, and performance spaces. 

Theater and music performances are primarily physical experiences that occur in dedicated venues, while many games spread out audiences and performers physically and virtually \cite{smuts2005video, crawford2007playing}. In conventional theatrical genres audience members are segregated from performers, with feedback curbed to pre-and-post performance clapping and cheering~\cite{lancaster1997spectators}, while more experimental experiences include showings of \textit{The Rocky Horror Picture Show},  encourage spontaneous audience participation~\cite{marshall1967medium}.  Less traditional theater allows audiences to contribute to the performance in a limited form, for instance employing a murder mystery audience to collectively choose a murderer \cite{schmitt1993castingAudience} or drive the dialogue~\cite{squinkifer2014coffee}. In contrast, games exist in a range of physical and virtual forms, from tabletop games like \textit{Dungeons and Dragons} that build narrative through a shared imaginative fantasy \cite{fine2002shared}, to video games that immerse players and audiences through graphics, animation, mechanics, and reward structures ~\cite{Tynansylvester2013designing, smuts2005video}. 


In juxtaposition to both theater and games, theme parks unify heterogeneous performances, games, and rides aimed at different audiences into an overarching group experience. Based on ancient and medieval religious festivals, trade fairs, and traditional amusement parks~\cite{milman200713}, themes parks integrate storytelling~\cite{bukatman1991there,schell2001designing}, simulation, and interactivity~\cite{shani2010role,milman200713} through primarily physical experiences. Together, the review of these domains aims to uncover additional insight that can inform iteration on a widely-applicable spectrum of interactivity.

\section{Characterizing Interactivity in Theater, Games and Theme Parks}
\label{sec:DomainLitReview}

The primary goal of this work was to gain insights about audience interactivity across diverse domains in order to build a common framework that could by practitioners in different domains. The following section relates insight from the codesign sessions to literature in three domains, theater, games, and theme parks, representatives of diverse audiences, interaction modalities, and performance spaces. 


\subsection{Dimensions of Audience Interactivity}

\subsubsection{Interactivity in Passive Experiences}
Traditional performances assume a clear distinction between the role of the audience and performers \cite{brooker2003conclusion}: audiences do not interact with performers or have a role in the direction of performance or narrative. Zimmerman ~\cite{zimmerman2004narrative} contradicts this assumption, suggesting that audiences can interact with experiences cognitively~\cite{zimmerman2004narrative}, as a psychological reader-response with contradictory and emotional interpretations, imbuing seemingly passive experiences as full of interaction. 

The literature also suggests that audiences participate in collective emotional experiences such as laughing or holding their breath, which validate their personal experiences; this helps explain why the presence of an audience is essential for a sense of `liveness'~\cite{reeves2005designing}. To understand and build on passive engagement, researchers and performers have employed several technologies to sense audience engagement, from simply watching audience expressions, to analyzing gestures and expressions using computer vision techniques~\cite{mcduff2015crowdsourcing}. 

In the codesign sessions described by Striner \cite{Striner2017_Transitioning}, sessions offered audiences the chance to opt out of interaction, wanted audience members to enjoy performances passively at times, and even decided that some music performances are not meant for interaction. This outcome suggests that passive experiences are as important as interaction, allowing audiences to absorb, appreciate, and reflect on the details of a performance, which is supported by psychology literature on interactive film~\cite{vorderer2001does}

\subsubsection{Reacting to other Audience Members}

Reacting to performance is a staple of the conventional audience experience \cite{lancaster1997spectators}, however 
literature suggests that audiences influence the experience of other audience members as they react to the experience.  Literature suggests that in theatrical experiences, audience members constructively play off each other during teamwork, conflict management, and negotiations \cite{peltonen2008s}. For instance,  an audience member could be influenced to give a standing ovation when others do. 

Theme park literature describes characterizes this phenomenon as a learning tool; for instance, at Wizarding World of Harry Potter, theme park goers watch other guests to learn the mechanics of `casting a spell' using an interactive wand \cite{burn2004potter, kronzek2010sorcerer}. This experience is further characterized by Reeves as an entertainment and teaching experience \cite{reeves2005designing} that allows audiences to study interaction while waiting their turn.

\subsubsection{Interactivity through Personalization}
Personalization in interactivity describes tailoring experiences to audience preferences, tastes, or capabilities. In Striner \cite{Striner2017_Transitioning}, codesign sessions expressed the need to acknowledge audience members' personal interaction preferences with the same sense of significance as fairness and democratization. Further, codesign groups also designed personal interactions for audience members; for instance, in session 1, a group designed an earbud-hat for audiences to contribute to a piano performance in a way that only they could hear.

Theme parks fully embrace personalized experiences in order to fully immerse audiences in fantastical worlds \cite{mccool2001should}; guests can meet characters \cite{ito2005intertextual}, and personally experience narratives \cite{schell2001designing} \cite{cross2005playful}; for instance, at the Wizarding World of Harry Potter, guests can be `chosen' by a wand at Olivander's wand shop, reminiscing over a scene from the first book in the series \cite{rowling1997harry}. Expert interviews further elaborated that audiences personalized experiences as they moved through space; for instance, a theater expert described that an outdoor interpretation of Swan Lake empowered audience members to ``\textit{become documentarian[s]\ldots start taking photos\ldots[and] climbing up on top of things.}'' Paralleling these physical experiences, recent advances in augmented and mixed reality technologies have likewise allow for games to be  personalized to the players location ~\cite{azad2016mixed, lv2015touch}, abilities ~\cite{togelius2011search, shaker2010towards}, and preferences \cite{summerville2016learning}. 

Audience interactivity codesign sessions \cite{Striner2017_Transitioning} also designed for `personalized extras,' clothing and prizes that allowed audiences to personalize performances for themselves, seeing value in `dressing up' for the show, a parallel to a performer putting on a costume. Using dress to personalize experiences is heavily paralleled in literature; Eicher's theory describes dressing up in fantasy costumes as a communication of the \textit{secret self}, where the bulk of fantasy interactions takes place ~\cite{eicher1981influence,fron2007playing}. Miller proposes a construct of fantastic socialization, where individuals play unrealized roles ``constructed only with the cooperative help\ldots and the contrasting foil provided by others" ~\cite{goffman2005interaction,miller1990critical}. Fron et al. define such personalization as a co-performative act with other spectators, gaining pleasure from the ingenuity and artistry that go into creating one's persona and costume ~\cite{bell1997ritual,fron2007playing,ito2005intertextual}. This style of personalization can be seen at contemporary American cultural festivals such as DragonCon~\cite{fron2007playing} and also reflects Zimmerman's ``meta-interactivity'' mode~\cite{zimmerman2004narrative}.
 
\subsubsection{Influencing Performers}

Primary examples of audience interactivity include audience members that influence or augment the performance experience without explicitly becoming a performer. Influencing performance includes visual voting systems~\cite{Unger_Photovote}, and audience input in improv~\cite{teambuildingThroughImprovisation}. While these types of interactions are popular, theater literature suggests that they are often asynchronous or inequitable~\cite{lancaster1997spectators}, prioritizing choices of audience members' closers to the stage~\cite{SarahHorncoverage_2013} or in positions of power~\cite{monson1990forced}.  Technical advancements have helped support democratic influence over voting. In an early example, audiences at 1967 World's Fair in Montreal voted on alternative endings to a film~\cite{anderson2008private}, and more more recent work in computer science has integrated real time audience interaction into algorithmic and computer-assisted musical composition~\cite{freeman2008extreme}.

Technical advancements in planning-based story generation~\cite{cavazza2002character} have helped support democratic influence over narrative; For instance, a play \textit{Coffee! A Misunderstanding}, allowed audiences to vote on dialogue that players read out during performances \cite{squinkifer2014coffee}. As well as influencing dialogue, Striner \cite{Striner2017_Transitioning} found that codesign groups considered how audiences could influence different sensory modalities, by control wind gusts to lift a bride's veil or tickle a performer with remote controlled feathers. 

\subsubsection{Augmenting the Experience}
As well influence the experience, literature suggested that audience members also wanted to augment the performance experience, such as by having audiences dance and move to a music experience~\cite{nettl1998course,strutner_2015}. For instance, Striner \cite{Striner2017_Transitioning} found that codesign groups designed tangible technologies, interactive hats, hand sensors, and palm pushbuttons that could interface with the musical performance. Groups transformed both music and feedback into experiences that could be manipulated with tangible technologies. For instance, one group suggested creating tangible `sound chips,' discrete bits of music that audiences could append to performance melodies.  

Research in computer music likewise suggests that audiences can augment performances by adopting a compositional role. Winkler notes that interactive music can ``\textit{create new musical relationships}'' between audience and performers ~\cite{winkler2001composing}; for instance, McAllister ~\cite{mcallister2004interactive} allowed audience members to add to a digital score synced to a real time display that musicians can read. Literature suggests that this compositional relationship between audience and performers can also be asynchronous; for instance, van Troyer \cite{vanTroyer_constellation} describes an interface for audiences to co-create asynchronously with composers, drawing `constellations'  that change the arrangement of
musical materials. Similar examples exist in interactive fiction design. For instance, Machado ~\cite{machado1999once} describes a storytelling environment, \textit{Once Upon A Time}, that develop characters, story themes and narratives out interactions with children.

\subsubsection{Bi-directional Influence}

Both physical and digital interactive performances lean heavily on the affordances of bi-directional interaction. For instance, gospel music uses call-and-response to nudge democratic audience participation~\cite{nelson1996sacrifice}, and computational narratives personalize player experiences by iteratively tracking and adapting narrative scheduling to player pacing ~\cite{azad2017linsp}. Similar research has produced a virtual dance partner that improvises dance moves based on audience actions ~\cite{jacob2015viewpoints}, and a narrative agent that respond to audience gestures with dialogue ~\cite{o2011knowledge}. 

As well responding to each other, some literature characterizes bidirectional interactions as `pushing and pulling' between audiences and performers. For instance Rickman~\cite{rickman2002dr} describes a text narrative system that uses word selection to reveal additional information about an object or action, which in turn drives narrative forward~\cite{cover2004interactivity}. Similar to this idea, codesign groups in Striner~\cite{Striner2017_Transitioning}, allowed performers to influence the type of food audience members had available during the show. Curiously, the research suggests that bi-directionality many not always be intentional. For instance, Van Maanen~\cite{van1991smile} describes how at Disney World, guests and cast members cyclically affect each other's emotions; cast-members are required to smile, however guests not smiling or waving back can ruin an operator's day. 

\subsubsection{Becoming Performers and Taking over Performance}

All three domains allowed audience members to take on performative roles, but differed in their approach. Games create immersion by giving players a sense of control \cite{cordova1996intrinsic},  allowing users to select strategies,  manage activities, and make decisions that affect outcomes \cite{shaker2010towards}. Video games have an inherent performative experience, allowing audiences to dually function as both players and audiences members \cite{smuts2005video}; playing and watching a game are different experiences, however literature suggests that games imbue players with spectatorship in between moments of play \cite{Taylor2010_SpectatorshipGames}. For instance, multiplayer LARPS (live-action role playing games) are considered a sort of performances-play experience~\cite{simkins2015arts}. Unlike a play, which requires a clear vision of context, characters, and relationship to be prepared prior to the show to have plausibility, since there are no official audience members, Simkins~\cite{simkins2015arts} describes how this work is done by the audience-performers, who use preparatory materials to give life the narrative. Fantasy sports games further blend the roles of audiences and performers~\cite{shipmanblending} by integrating the ``activity in a virtual game and spectatorship of a real sport;'' players act as `coaches,' selecting players from a sport to be part of their team, and get points based on players' real performance~\cite{shipmanblending}. Developments in large-scale streaming, tangible interfaces, and virtual and augmented reality have fundamentally changed  the game viewer landscape; Twitch streaming allows audiences to watch, comment on, and interact with streamers during games~\cite{toupsHammer_2014Gamesframework}, and augmented reality has given players and viewers a way to interact in a physical space, such as a Harry Potter augmented reality experience that allowed users to experience a Harry Potter narrative~\cite{HArryPotterAR_gupta_shah_george_pramer}.

Although less accessible than games~\cite{cross2005playful}, theme parks fully embrace audience in performative roles, integrating storytelling \cite{bukatman1991there}\cite{schell2001designing}, simulation, and interactivity \cite{shani2010role}\cite{milman200713}, and emphasizing
physical experiences. Theme park experiences often give audiences a chance to re-experience character roles and narratives; for instance, at Universal's Wizarding World of Harry Potter, allows guests to eat the food Harry Potter ~\cite{rowling1997harry} ate at the Leaky Cauldron~\cite{fowler_2016_HarryPotterFood}, `cast spells' using interactive RFID wands~\cite{HarryPotter_Wands}, and to live through a fight scene on the Hogswarts Express train train~\cite{HarryPotter_HogwartsExpress_themeparkworldwide_2014}.
These firsthand narratives lean heavily on multisensory, spatial and temporal experiences~\cite{milman200713} to give a sense of presence \cite{palmer2010parenting}~\cite{bukatman1991there}. Theme parks also give audiences the chance to take control of performance interaction; at Disney World, guests in line for a Peter Pan ride can play with an interactive shadow puppet displays that let guests make music with bells, play with butterflies, and even decide to release Tinker Bell from being trapped in a lantern~\cite{PeterPan_ShadowPuppets_andersson_brigante_2015,BestQueuesDisney}. Performative roles are so integral to theme park experiences that experiences must be designed for a wide range of audience~\cite{mccool2001should} that might not necessarily understand game metaphors like cut scenes, life meters, and levels, and must entertain guests in failure as well as success~\cite{schell2001designing}.

In theater, the role of audiences as a performative agent is contested. In Hamlet on the Holodeck, Murray~\cite{murray2017hamlet} suggests that audience participation may be `awkward' and potentially `destructive'; she describes a Woody Allen story, the \textit{Kugelmass Episode}~\cite{WoodyAllen_KugelMass} where a humanities professor is given the opportunity to jump into the pages of Madame Bovary, only to confuse the narrative of the novel; `Who is this character on page 100? A bald Jew is Kissing Mme Bovary?'' With this, Murray points out that "when we enter the enchanted world as our actual selves, we risk draining it of its delicious otherness."~\cite{murray2017hamlet}. Instead of becoming a maintaining a traditional performer role, many theatrical music experiences allow audiences to participate in social co-creation to help audiences make sense of and appreciate complex arts~\cite{nelson1996sacrifice}. For instance, Whitacre~\cite{whitacre} developed a virtual choir that allowed singers all over the world to contribute to a performance, and Machover created \textit{City Symphonies}~\cite{hoffman2015design_MachoverCitySymphony} to create symphonies unique to different cities. Performative theatrical and music experiences have also been designed to build self-esteem ~\cite{nettl1998course}. For instance, Boal created the \textit{Theater of the Oppressed} to promote   social and political change; in this medium, audience members became `spect-actors,' who used the medium to explore, show and analyze their experiences. Likewise, at home music experiences like Guitar Hero~\cite{bernardo2014music} and Hyperscore~\cite{farbood2004hyperscore} helped bridge  skill gaps.

\section{Expert Interviews}

\subsection{Method}
We conducted 8 in-depth interviews with interactivity experts in the domains of theater and music, games, and theme parks. Experts were researchers and practitioners with a minimum of 3 and an average of 13 years of professional or academic experience working on interactivity in their respective domain. 

Each interview lasted approximately one hour. Experts were asked to describe their background in interactivity, themes and trends they had encountered, challenges in interactivity, and the role of technology in interactivity in their domain. The interviews concluded by asking experts for explicit feedback on prior work detailing a spectrum of audience interactivity as described by Striner \cite{Striner2017_Transitioning}, whether the experts thought the spectrum described interactivity based on their experiences, and what elements the spectrum not capture well. During the interviews, an interviewer transcribed dialogue in a shorthand form and took notes. After the interviews, missing dialogue was transcribed from interview audio recordings.  

After the interviews, all interview transcripts were compiled into a unified Google Doc. Three coders thematically color-coded important themes in personal versions of the document, using qualitative open coding methods ~\cite{strauss1990open}. After this, coders grouped themed quotes together to understand the relationships and commonalities between the themes. The process took several stages of iteration; each coder moved quotes around and renamed themes until they felt that each theme most effectively captured the interview's data and intention. After individually iterating on themes, coders discussed and clarified their thematic groups, then merged all overlapping themes together into the unified transcript document. In the annotated unified transcript, coders discussed parallel theme nuances and identified a final set of themes and sub-themes.  

\subsection{Themes}
\subsubsection{Expressing, Employing and Shaping Culture}
A primary theme in the expert interviews was how interactivity design influenced and was influenced by cultural settings and norms. Experts described a range of social experiences, from deeply personal experiences that "\textit{wanted to subvert the...generally flippant...treatment of death in video games}" to those steeped in cultural norms; for instance, a theater designer characterized eating at a restaurant as a "\textit{performative...ritual...you get seated, get a menu...[and follow] a script for ...[getting] served.}" Likewise, a theme park designer identified weddings as a \textit{"culturally sanctioned form of LARP...[full of] expectations...multisensory event design...[and] interactive performance.}" Employing the social mechanics of weddings,  
the designer further described an experience she had designed to nudge wedding attendees to socialize and get to know each other. Culture and ritual were central to effective design; she underscored the importance of "\textit{not [undermining]}" the overarching experience; the interactive experience she built had to work "\textit{within [the bride's] vision for the day.}" 

\subsubsection{Experimenting with Design}
Design culture was equally important in experience design.  In games, one expert explained that the ``\textit{[design] process shapes what gets made};'' large game companies with resources to build high definition experiences have difficulty producing unique mechanics because the work involved creating them cuts cross across job expertise. In contrast, the rise of the individualistic maker culture has allowed designers to experiment with novel modalities, experiences that don't ``\textit{look like a computer game}.'' In our interviews, one expert described building custom controllers for a game, another described building a game played on an embroidery machine, and a third described constructing emergent narratives using artificial intelligence algorithms. Likewise, experts expressed interest in developing avant-garde experiences; a theme park expert had experimented with constraints, describing a game mechanic that relied on `\textit{`coping with limited bandwidths of information.}'' Similarly, a theater expert had experimented with an ``\textit{augmented radio play,}'' where a larger audience watched and directed an individual’s experience.

This culture of experimentation also extended to choice of settings. Rather than building experiences for profit, a game and theater designer explained that many interactive experiences were built by enthusiasts, researchers, and students for festivals or conferences. Venue could fundamentally change an audience's experience; in line with Rouse’s description of early films as carnival spectacles ~\cite{rouse2016media}, this expert also described the importance of designing the external `facade' of the show; people in line had a chance to peek into the interactivity `wizarding' station, getting insight into the back-end mechanics after having the experience.  Although unique, a game expert acknowledged that these experimental experiences were limited to audiences who could be physically present at performances. 

Although experimentation was at the forefront of interactivity, several experts noted the importance of scalability and robustness. A theater expert noted that scalability was important because it ``\textit{enables resource intensive [experiences}],'' and other experts affirmed this, commenting that scalability was ``\textit{more practical}'' since many interactive experiences have a  ``\textit{cost to upkeep and maintain technical components}.'' To create scalable experiences, a theme park expert remarked that interactions had to be ``\textit{incredibly robust}'', sometimes having ``\textit{having two possible paths [or] tracking people so that they don't collide [while] managing the performance and actors}.  The theme park designer summarized the challenges of building technology for an Indiana Jones interactive line experience; enormous challenges existed in building minimally engaging interactions in an unsupervised space that were committed to the theme of the narrative,  ``\textit{you need to...design a fun physical experience...that people can’t break,}'' and faced with logistical considerations,``\textit{interactivity [often became] secondary}.''

\subsubsection{Designing Interactions that Transcend Novelty}
Technology was foundational to most interactions experiences, but experts emphasized the importance of designing substantive interactions that transcended novelty. One theater expert suggested that productions often incorporate technology as an novel afterthought; ``\textit{immersive theater and interactivity are buzzwords,}'' he explained, ``\textit{so the easiest thing for a theater company to do is to throw a bunch of iPads a wall.}''  The expert further described that ``\textit{[iPads work best]...if creating performances in virtual worlds,}'' but ``\textit{if [you] put iPads in a Victorian piece, there’s a mismatch [in the experience].}'' Similarly, another expert noted that ``\textit{shockingly few movies benefit from 3D},'' explaining that ``\textit{there should be a reason for technological choices.}'' Although intention was key, experts also balanced intentionality with practicality. One expert explained the practicality of `` \textit{hooking into [the] certain level of tech that’s already on the scene},'' citing that ``\textit{a lot of scavenger hunt experiences [use] participants’ own smartphones.}''

In order to artfully choose technology, experts described the importance of examining the role of interactivity in the narrative; one expert explained that ``\textit{interactivity [should be] something that the artist considers, rather than an element of production.}'' Further, interactivity choices had to deliberate and explicit; multiple experts noted that creators needed to understand ``\textit{why you want something to be interactive...and when it [has] to be interactive.}'' For instance, a theater designer described an opera, Lilith~\cite{Lilith_uctelevision_2015}, that integrated technology and aesthetics into using mirrored wall displays that created the impression of movement through time and space.

\subsubsection{Control and Uncertainty}
The logistics of control and chance were a primary concern for experts. Types of audience interactions and extent of audience agency and control were weighed against throughput; a games expert described the rise of flash games with limited interactivity and ``more noticeable interaction outside the artifact...a social ecosystem...[of] comments...and ratings." This ``\textit{software on the small}'' has become popular ``\textit{due to the ``barrier [being] incredibly low.}'' These limited interaction experience worked well for large communities; for instance, a game designer described that Twitch streaming communities can have many people can participate ``\textit{asynchronously or asymmetrically,''} but designers had to set the interaction bar low in order to ``\textit{get a thousand people together to do [one] thing}.''  In contrast to low-bar experiences, another expert described quite the opposite with an in-depth `high attention' transmedia experience experienced by a oneperson at a time; In order to "\textit{let experience breathe...[allowing players to be] more forthcoming,}" they had to adjust the length of the experience, from 15 to 45 minutes. In order for the audience to have significant control of the experience, designers had to limit the audience to a single person, and triple it's length of time.

Chance balanced this need for control. Several experts discussed how human behavior is unpredictable, and allowing for interaction increased the uncertainty in the performance. A games expert noted that ``\textit{when you are working with audience...there is chance...something you cannot count on,}' and a theater expert explained that `` \textit{the value of going to the theater...[is] you have this variability...that something could go wrong, something unexpected.}'' Because of this, experts explained that designers could not count on a narrative to progress in a particular way. For instance, one expert told a story of a military-trained participant who unintentionally subverted their interaction experience by refusing to adopt a cover story integral to his character's interaction. To address this uncertainty in design, an expert described procedural content generation algorithms~\cite{azad2016mixed} that create an emergent narratives using simulation and  theatrical elements~\cite{squinkifer2014coffee}. Although there was no way to predict how the player might respond, procedurally generated content could help nudge participants toward appropriate interactions.

\subsubsection{Considering a Contract of Care}
Expert interviews suggested that complex audience interactions have an implicit `\textit{`contract of care};'' when audience members are asked to perform, interactivity should consider ``\textit{how willing ...and prepared [audience members are] to interact, and prevent them from feeling pressured or secluded.}'' For instance, a designer described that a transmedia experience they build felt so "\textit{real and emotionally intense,}" that players would be "\textit{crying at the end,}" and a game designer described needing to "\textit{safely prevent emotional spillover in LARPS...}", such as separating game and real world "\textit{romance [elements].}"

Intense experiences required onboarding that "\textit{cut the playing area off from the rest of the world...[and] separated [the] player from [the] act of role-play.}" For instance, in the aforementioned transmedia experience, audience members were asked to adopt a cover story. A games and narrative expert explained that playing role in which the character has to impersonate a second character insulates players from embarrassment because players could attribute mistakes to their character, not themselves. 

\subsubsection{Multiple Dimensions of Interactivity}
Through the discussions, several experts commented on the fluidity and range of interactivity that existed in their domains. Multiple experts, for instance, described seemingly passive experiences as full of interaction; audience members watching a performance choose to focus on a particular part of a stage, or participating in  collective reactions, like laughing or holding your breath, that helps validate our personal experiences. Experts attributed this range of interaction to various factors associated with the experience. Two experts (one theater and one game designer) even described the many types of interactions as `a continuum;' the theater expert described ``\textit{much agency that the performer grants an audience}'' whereas the game designer described a continuum of ``\textit{directorial roles,}'' in line with Samuel’s work ~\cite{samuel2016computatrum}.

\begin{enumerate}
\item \textit{\textbf{Agency}}
Several experts alluded to the ``\textit{sense...or the illusion of agency}'' during interviews. A theater expert differentiated between having a``\textit{physical presence on stage and the autonomy of agency;}'' physical presence, they explained, was not necessarily interactive, whereas audiences could ``\textit{affect outcome or dialogue}'' without physically controlling the experience. Although agency was described differently by experts, and was subject to distinct constraints, it was integral to the experience. One expert emphasized that audience members having ``\textit{any kind of agency through technological elements.... can be very awe inspiring.}'' 

\item \textit{\textbf{Space}}
Experts suggested that the affordances of  venues and spaces affected the experience of individual audience members.  A theater expert explained that ``\textit{as an audience member...things are happening [all] around you...you have to decide to look...`edit’ your perspective of that experience.}'' Movement through space, for instance, gave agency to audience members; the theater expert described how an outdoor interpretation of Swan Lake empowered audience members to  `` \textit{become documentarian[s]\ldots{} taking photos, [and] climbing up on top of things\ldots{} not just observing, because  the ``rules [of the performance] have changed}'' for them. This supports literature suggesting that space and proximity affects interaction experiences~\cite{michelis2011audience, peltonen2008s}.

\item \textit{\textbf{Multimodality}}
Multisensory experiences were a strong theme present in a majority of the expert interviews.  Experts suggested the importance of  stimulating multiple senses, explaining that audiences ``\textit{itch [to experience] other senses}.'' One expert, for instance, compared having a limited sensory experience to watching a color movie in back and white. Using tools like Arduino, experts suggested blending physical and virtual storytelling elements to create fulfilling experience dynamics. For instance, one expert detailed how a narrative generation virtual game he created ``\textit{never took off}'', but when he integrated virtual storytelling into a ``\textit{weird hybrid...it was way cooler than the digital game.}'' Another expert noted the `\textit{magical}' quality of physicality into experiences lent a ``\textit{human element...to disrupt the coldness of purely digital experience.}''

\end{enumerate}

\section{Discussion}
The goal of this work is to develop a framework to describe audience interactivity across a broad range of experiences. Building on early work by Striner~\cite{Striner2017_Transitioning} which organically proposed a spectrum of interactivity for children, this work developed a framework and vocabulary for researchers, designers, and artists in entertainment interactivity to compare and consider divergent interactivity design. To develop this framework, we surveyed literature across three entertainment domains, theater, games, and theme parks, then interviewed eight experts in those domains. The discussion overviews emergent patterns and commonalities in the literature and expert interviews that informed our design.

\subsection{Levels of Audience Interactivity}
A primary finding of our work substantiated the presence of a spectrum that characterized interactivity based on agency. Several experts independently described the presence of a `\textit{a continuum};' in which audience members had `\textit{range of agency},' but when asked for directed feedback about the spectrum described by Striner~\cite{Striner2017_Transitioning}, experts in the different domains also debated its range; a theater expert and composer felt that most audience interactivity would fall on the passive end of the spectrum because interactive moments ``\textit{were dangerous\ldots to control},'' whereas a game designer suggested the opposite, that interactivity in games ``\textit{starts at 4 (audience influencing performers)\ldots [and] clusters around the right [end of the spectrum]\ldots everything else doesn't seem relevant.}'' Although some experts didn't understand the extent of the spectrum, others valued its broad range. For instance, a game designer commented that having less interactivity for game audiences made sense as a design choice; ``\textit{audiences are big\ldots [having large audiences in player roles would] become an incoherent mess\ldots you can't\ldots scale up.}'' Likewise, an composer responded that on the ``\textit{chart\ldots every [level] could be done [successfully]\ldots if it makes sense [artistically]}.''

The literature review affirmed the presence of this continuum, finding that interactivity ranged from passive to active experiences delineated by agency of individual audience members; passive and personalized experiences gave audiences agency over themselves, and reactions, influencing, augmenting, and becoming performers gave audiences  agency over other audience members, performers, and the overarching experience. Although the literature supports the presence of different levels, we found that interactivity was more prominent in some domains; for instance, theater and music predominantly uses interactivity to influence and augment performances~\cite{winkler2001composing,squinkifer2014coffee,vanTroyerSleepNoMore}, games employ audiences as performers~\cite{shaker2010towards,mateas2006interaction}, and theme parks focus on experiences full of personalization and bidirectional influence~\cite{van1991smile,schell2001designing}.

\subsection{Parallel Dimensions of Interactivity}
Striner~\cite{Striner2017_Transitioning} described a spectrum of interactivity, however this work found several dimensions of interactivity paralleled this spectrum. 
This was substantiated by experts' directed feedback about the spectrum; during interviews, several experts questioned the meaning of interactivity, asking whether it referenced to ``\textit{physical presence on stage versus autonomy or agency?}'' or control over the ``\textit{explicit shared experience.}'' A theater designer posited that interactivity could have multiple dimensions, such as control over the narrative, design, linearity of experience, or control over specific elements of design~\cite{everett1986communication, laurel1986interface, goertz1995interaktiv, brignull2003enticing, reeves2005designing}. 

In addition to audience agency described by the levels of interactivity, the expert interviews found that parallel dimensions of space, control, and interactivity moderated audience members sense of agency and affected the overall interaction experience.  Together, these parallel dimension can help to inform several expert interview themes; how to effectively express and shape culture, to balance experimentation with practical concerns to design meaningful experiences that transcend novelty and make audiences feel comfortable with interaction.



%

\begin{figure}
  \centering
  \includegraphics[width=.75\columnwidth]{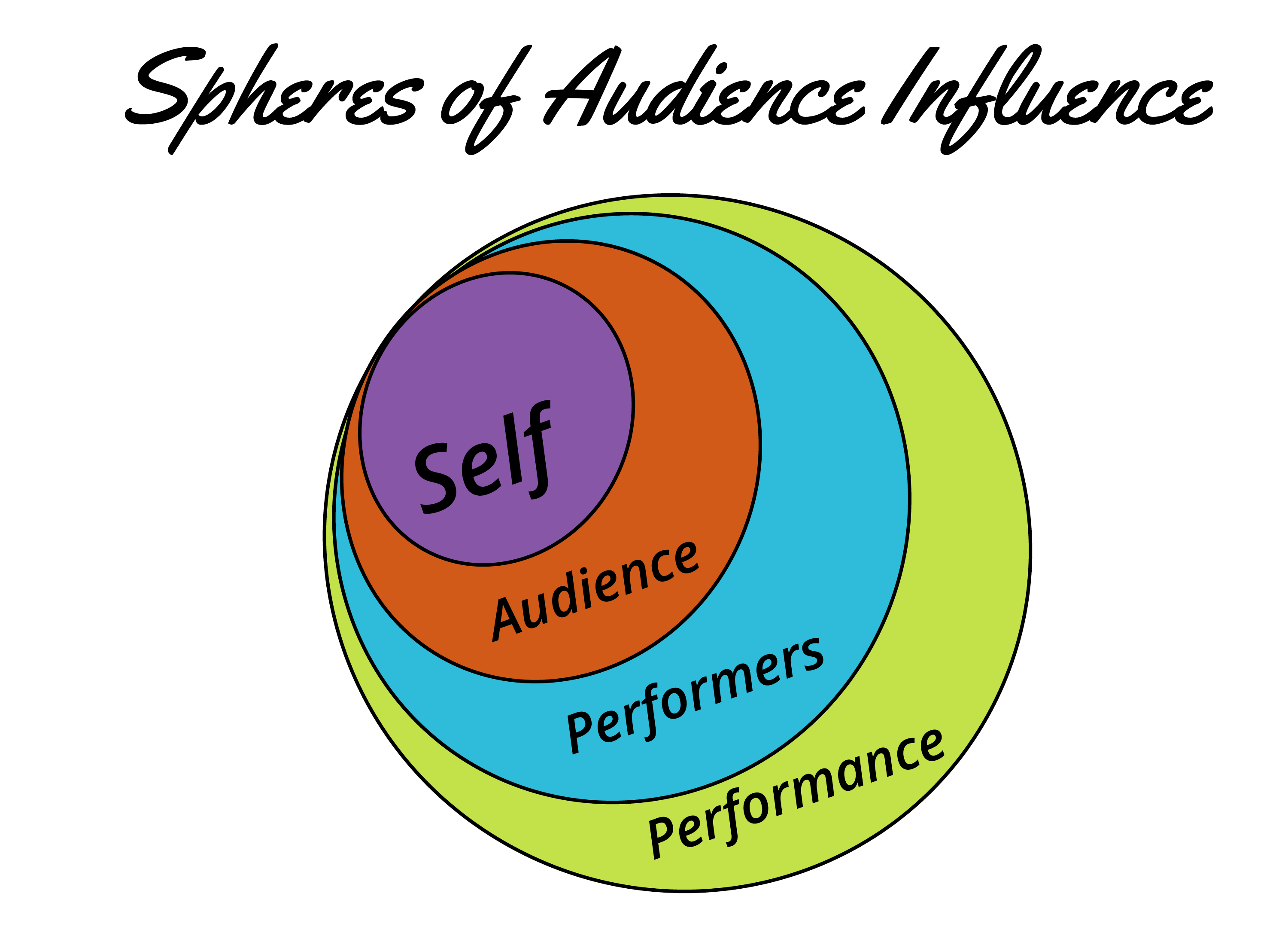}
  \caption{The audience spheres of influence used to describe the spectrum of audience interactivity. The innermost sphere of influence is over yourself, other audience members, performers, and finally the performance.}~\label{fig:SpheresInfluence}
\end{figure}

\begin{figure*}
  \centering
  \includegraphics[width=\textwidth,height=6.5 cm]{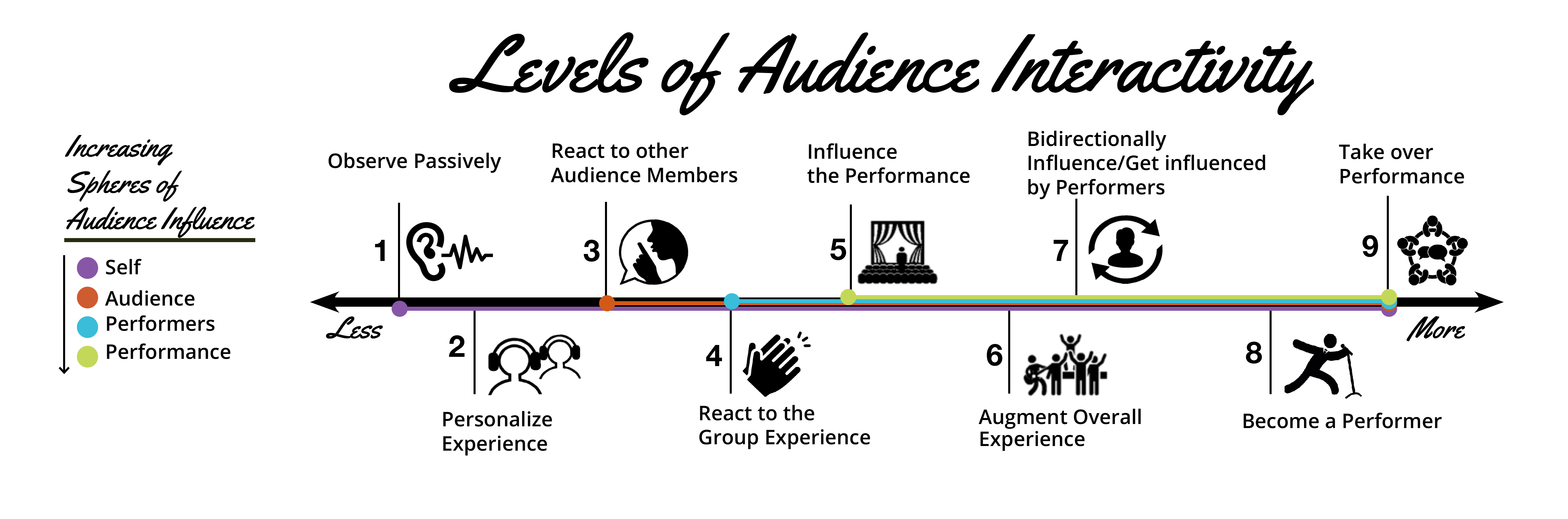}
  \caption{Levels of of audience interactivity mapped to the audience spheres of influence. Interactivity levels include audience members passively observing performance, personalizing performance, reacting to other audience members, reacting to the performance, influencing performers, augmenting the performance, audience and performers bidirectionally influencing each other, audience members becoming performers, and finally taking over the performance}~\label{fig:SpectrumDimensions}
\end{figure*}

\section{Proposed Audience Interactivity Framework}
The literature review and expert interviews informed the design of the audience interactivity framework, which comprises a spheres of audience influence,  The framework comprises  new spectrum of audience interactivity. Shown in figure \ref{fig:SpectrumDimensions}, the new spectrum introduces an framework for audience spheres of influence, presents the new spectrum mapped to the spheres of influence, and characterizes interactivity dimensions orthogonal to the spectrum. 

\subsection{Spheres of Audience Influence}
The experts' divergent interpretations of interactivity provoked us to develop a framework with which to define the new spectrum. This framework, the \textit{Spheres of Audience Influence} is presented in figure \ref{fig:SpheresInfluence}. Audiences are the focus of the spectrum, so the framework defines interactivity by the extent to which an audience member affects others around him; in the innermost circle, audience members influence their own experience, in the second circle they influence other audience members, in the third circle they influence performers, and in the outermost circle, they influence the larger performance. The spheres are nested, so each circle adds to an individual audience member's layer of influence. In this way, an audience member influencing the performance also influences performers, other audience members and themselves.

\begin{figure*}[t]
  \centering
  \includegraphics[width=\textwidth,height=8.5 cm]{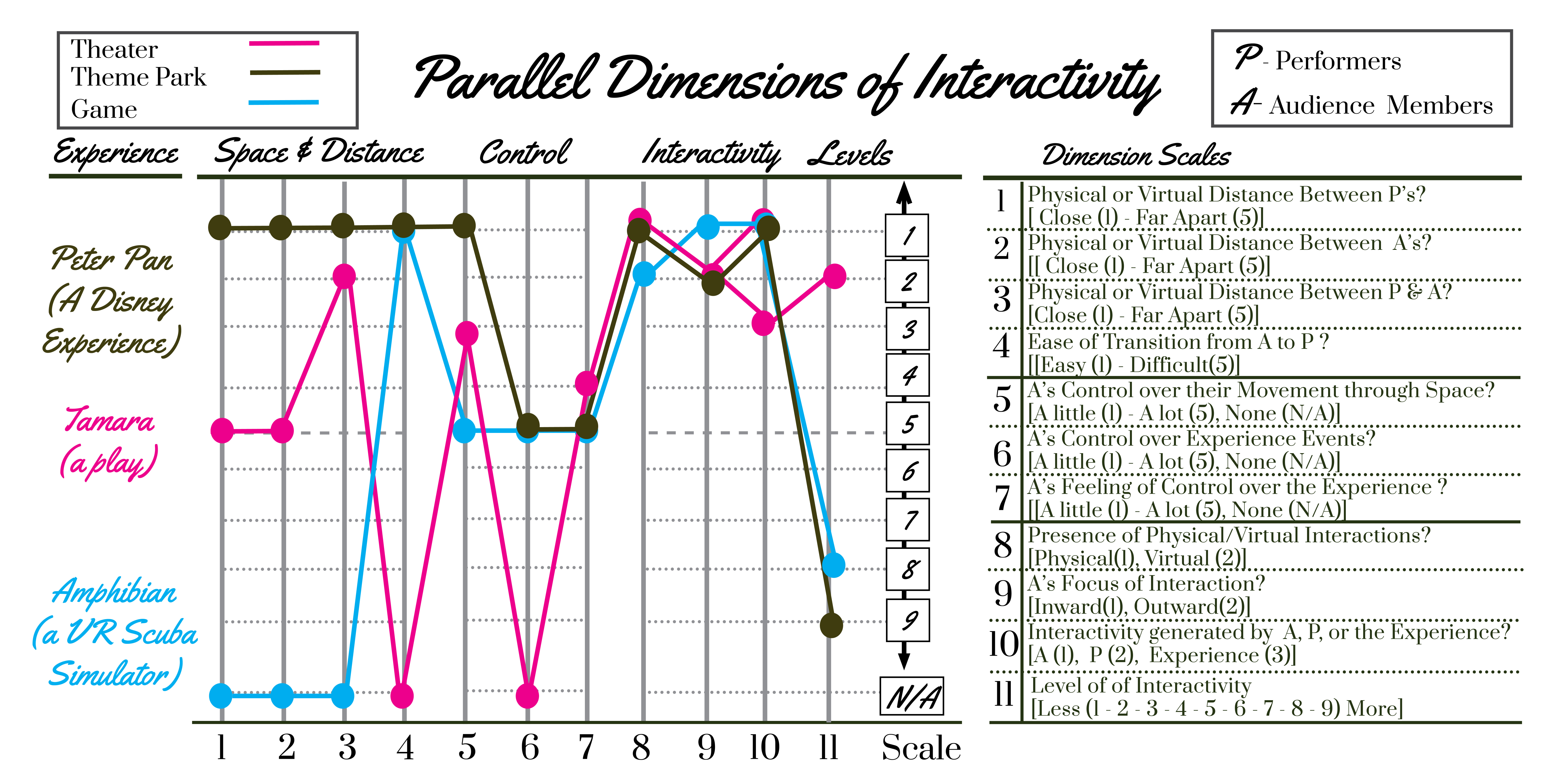}
  \caption{The parallel spectrum dimensions mapped to a Tamara,a theater experience, Amphibian a simulator game, and Peter Pan, a theme park experience.}~\label{fig:OrthogonalDimensionLines}
\end{figure*}

\subsection{Levels Of Interactivity}
Presented in figure \ref{fig:SpectrumDimensions}, the new \textit{Levels of Interactivity}, mapped to the \textit{spheres of influence}, expand on Striner's \textit{Spectrum of Interactivity }~\cite{Striner2017_Transitioning} using findings from the literature and expert interviews. Least interactive on the spectrum are 1) \textit{observing passively}, referring to an audience member cognitively shaping their experience, and  2) \textit{Personalizing their experience}. More interactive is 3)\textit{reacting to other audience members}, allowing audience members influence one another, such as responding to a comment on YouTube. In 4) audience members \textit{react to the performance}, influencing performers, and in 5) \textit{influencing the performance}, they exert indirect control over the overall experience. For example, virtual audiences watching a Twitch stream could suggest a way for a streamer to solve a puzzle. In contrast to 5, audience members in 6) \textit{augment the overall performance experience} without explicitly becoming performers, such as by deciding to dance to music at a rock concert. In 7) \textit{bidirectional influence} between audience and performers, performers explicitly respond back to audience's influence or reactions, such as Mickey Mouse waving back to a child at Disney World. Levels 8) \textit{audience members become performers} and 9) \textit{audience members take over the performance} both give audience members an explicit role in the performance but in the former, performers are in control, and in the latter, audiences take control of the performance. Audience members sing along with a choir would become performers, whereas an audience member invited to perform karaoke would take over the performance. 

\subsection{Parallel Dimensions of Interactivity}
The levels of interactivity fit into a set of parallel dimensions of interactivity shown in figure \ref{fig:OrthogonalDimensionLines}. In order to fully describe the different dimensions of interactivity, we developed a set of orthogonal dimensions that complement the spectrum of interactivity. These dimensions lie on a parallel coordinate chart, which is commonly used to visualize multivariate data with high dimensional geometry to find patterns in the data~\cite{inselberg1985plane,gemignani_ParallelCoordiante}. To show a pattern in a set of data elements with n-dimensions, the chart typically contains n lines, typically vertical and equally spaced out. The data element is represented by polylines or connected points with vertices on the parallel axes. 

The parallel dimensions are presented together so that interactive experiences may be mapped onto one another; future work plans to survey a broad range of experiences to look for patterns in interactivity with the goal of practitioners using them to design interactive experiences. Three examples are shown mapped to the parallel dimension chart: Tamara~\cite{TamaraPlay}, an interactive play where audience members chose to follow different groups of performers through divergent narratives, Amphibian~\cite{jain2016immersive}, a scuba diving VR game experience, and Peter Pan~\cite{PeterPan_ShadowPuppets_andersson_brigante_2015}, an interactive shadow puppet experience audience participated in while in line for the Peter Pan ride at Disney World. Mappings were performed by coauthors, who experienced these interactive experiences firsthand. Each experience was mapped simply to preserve clarity, since experiences may be complex and composed of several different interactions, all dimension scales could be mapped to two or more choices.

The orthogonal dimensions are grouped by themes that emerged from spectrum findings, space, control, and interactivity. The \textit{space and distance} category describes the distance between performers (1) distance between audiences members (2) distance between performers and audiences (3), the difficulty of transitioning from audience member to performer (4). In Tamara, performers and audience were relatively far apart from other members of their group, but were relatively close to each other. In contrast, during the Peter Pan experience, performers and audience members were all located close to each other. Amphibian was creator for one person, so the relative distance between performers and audience members was not applicable. In Peter Pan and Amphibian it was easy to transition from audience to performer, however it was not possible to become a performer in Tamara.
 
Control was likewise an important theme that emerged from the literature and expert interviews; the \textit{control} category describe audience members control over their movements through the space (5), over experience events (6), and their overall feelings of control (7), which may not correspond to actual control over the experience. All eight of these dimensions lie on a Likert scale of 1-5 with an N/A option. Elements of control varied significantly across the three examples. Peter Pan audiences had little control over their movement in space, but had and felt a lot of control of the experience. In Amphibian, audience members had a lot of control over movements and events, and felt in control. In contrast, Tamara audiences had some control over the movements and felt relatively in control, but had no control over experience events.

The \textit{interactivity} category characterizes the parallel dimensions of interactivity. Physical or virtual modality of interactions (8) indicates the presence of (1-physical) and/or (2-virtual) interactions, and the focus of the audience's interactivity (9) indicates the focus of interaction as inward (1-focus on the self) or outward (2-focus on the experience). Dimension (10) describes the origin of interactivity  (whether interactivity was generated by 1-audience members, 2-performers or 3-the overall experience). Peter Pan and Tamara interactions were primary physical, focused outward, and generated interactivity by audience members. In contrast, Amphibian was a primarily virtual experience where interaction was generated by audience members and focused inward. 

Finally, dimension (11), maps the levels of interactivity described in figure~\ref{fig:SpectrumDimensions} to a 9-level scale. Interactivity in Tamara was primarily characterized by personalization of experience (level 2) because audience members chose which group of performers to follow. In Peter Pan, audience members primarily became performers (level 8), creating and interacting with a high level narrative as they played with shadow puppets. Amphibian was an exploratory underwater scuba experience that individuals had full control of, corresponding to level 9, taking over the performance experience. 




\section{Conclusion}

The goal of this work was to develop a framework to was to explicitly characterize the many ways in which audiences can interact with  experience across a range of performative domains. The framework aims to be a useful resource for researchers, designers, and artists in entertainment interactivity to consider opportunities for interactivity. While the framework aspires to be comprehensive, edge cases that do not fully map to the spectrum undoubtedly exist. Since new tools and new media continually reshape the interactivity landscape, we consider such cases to be good fodder for discussion, allowing us to better understand new forms of interactivity. 

Multiple themes emerged from our research that should be considered by practitioners, explicitly considering interaction goals, being aware of audience abilities and emotional needs, and considering a diverse range of interactions. We believe that our audience interactivity framework will allow practitioners from diverse fields to learn from one another.

This research presents a redesign of the spectrum of interactivity, however this is an early effect to characterize the many dimensions of audience interactivity. The spectrum does not endeavor to describe audience interactivity from the perspective of the performer, which may have unique interactivity characteristics and dimensions, or describe audience characteristics (e.g. culture, size and location), although they may affect interactivity.

\subsection{Future Work}

Future work will validate the clarity, precision, and effectiveness of the spectrum and orthogonal dimensions with audiences and experts in a range of domains. In order to help practitioners learn from other domains, we plan to use the framework described to survey a range of audiences, performers and creators of interactive audience experiences. Since people do not make consistent qualitative judgments \cite{seale1999quality}, we will also test inter-rater reliability within performances. 

This future work will allow designers to compare diverse interaction experiences and identify patterns that emerge across domains. This will enable designers to actively consider the novelty and practicality of their interactivity designs; identifying patterns 
of interactivity will help designers anticipate challenges that may arise in unconventional designs and nudge designers to experiment with interactivity.

\section{Acknowledgements}
\bibliographystyle{SIGCHI-Reference-Format}
\bibliography{sample}

\end{document}